\documentclass[11pt]{article}
\usepackage{graphicx}

\newsavebox{\hflrar}
\sbox{\hflrar}{\makebox[0pt][l]
{${\scriptstyle \leftharpoonup}$}{${\scriptstyle \rightharpoonup}$}}

\def \to {\rightarrow}

\begin{document}
\begin{center}
{\Large\bf On Transverse-Momentum Dependent Light-Cone Wave
Functions of Light Mesons} \vskip 10mm
J.P. Ma and Q. Wang    \\
{\small {\it Institute of Theoretical Physics, Academia Sinica,
Beijing 100080, China }} \\
\end{center}

\vskip 1cm
\begin{abstract}
Transverse-momentum dependent (TMD) light-cone wave functions of a
light meson are important ingredients in the TMD QCD
factorization of exclusive processes. This factorization allows one
conveniently resum Sudakov logarithms appearing in collinear
factorization. The TMD light-cone wave functions are not simply
related to the standard light-cone wave functions in collinear
factorization by integrating them over the transverse momentum. We
explore relations between TMD light-cone wave functions and those in
the collinear factorization. Two factorized relations can be found.
One is helpful for constructing models for TMD light-cone wave
functions, and the other can be used for resummation. These
relations will be useful to establish a link between two types of
factorization.

\vskip 5mm \noindent
\end{abstract}
\vskip 1cm
\par\vfil
\eject In the collinear QCD factorization for exclusive
processes\cite{BL,CZrep}, such as the form factor of $\pi$, the
nonperturbative effects are included in various light-cone wave
functions. At leading twist, transverse momenta of partons entering
a hard scattering are neglected. The hard scattering can be studied
in perturbative QCD. In this approach the perturbative expansion for
the hard scattering often has large corrections, in the form of
large Sudakov double logarithms, around the end-point regions in
which one parton in a hadron carries almost all the momentum of the
hadron. These large corrections will make perturbative expansions
diverge and a resummation is needed for them.
\par
The solution for resumming these corrections is suggested in
\cite{BS,LB} by taking transverse momenta of partons into account,
in which one introduces transverse momentum dependent (TMD)
light-cone wave functions similar to the light-cone wave functions
in the collinear factorization. We will call the latter as the
standard light-cone wave functions.
Before a detailed discussion of these wave functions some explanation is needed
for the nomenclature. The light-cone wave functions in the collinear factorization
here are called quark distribution amplitudes in the pioneer work in \cite{BL}.
The transverse momentum dependent light-cone wave functions were also
introduced in the light-front Hamiltonian formulation of QCD and called
as wave functions(see the review \cite{BPP}). However the TMD light-cone wave
functions defined in this letter are slightly different than those wave functions, the difference
is because of light singularities and will be discussed later.
With the TMD
light-cone wave functions one can make a TMD factorization instead
of the collinear one and show that the Sudakov logarithms can be
resummed. Because the TMD light-cone wave functions are
generalizations of the standard ones, one may expect to obtain the
standard light-cone wave functions from the TMD ones simply by
integrating the transverse momenta. However, this is not the case.
In this letter we explore in detail the relationship between these
two types of wave functions.
\par
Similar problems also appear in the collinear factorization for
inclusive processes like semi-inclusive DIS and Drell-Yan process as
well as exclusive $B$-decays, where large corrections appear around
edges of kinematical regions. In inclusive processes like Drell-Yan
and semi DIS one can introduce TMD parton distributions and a TMD
factorization for the differential cross-sections can be
obtained\cite{CS,CSS,JMY}. In this formalism one finds that the
resummation can be conveniently performed and there is a factorized
relation between TMD parton distributions and the distributions
appearing in the collinear factorization. In exclusive B-decays,
where $b$-quark is described by heavy quark effective theory, the
TMD light-cone wave function of $B$-meson can also be consistently
defined\cite{MW1, LiLiao} and its relation to the light-cone wave
function in the collinear factorization is explored in \cite{MW1}.
With the TMD light-cone wave function one can show that the TMD
factorization for radiative leptonic decay of $B$-meson can be
verified at one-loop level and the large correction around the
end-point region can be resummed\cite{MW2}. Using wave functions
or TMD light-cone wave functions
the form factor of $B\to \pi$ transition can also
be formulated in a factorized form\cite{BH}.
In other $B$-decays it
has been shown that the Sudakov logarithms can be resummed (see
e.g., \cite{Bsum} and references therein). It should be pointed out
that TMD factorization is not only helpful in resumming large
logarithmic corrections but also in probing the 3-dimensional
structure of hadrons. Various physical quantities can be
represented with these wave functions, e.g., various form factors\cite{BHMS},
generalized parton distributions\cite{DFJK}, single spin asymmetries\cite{SSA}, etc.
\par
TMD light-cone wave functions can be defined from the matrix
elements of quark and gluon operators in QCD. They have been
classified in terms of partonic configurations for various
hadrons \cite{JMYWF}. In the letter we will take $\pi$ as an
example, although it is straightforward to extend our results to
other hadrons. We will use the light-cone coordinate system, in
which a vector $a^\mu$ is expressed as $a^\mu = (a^+, a^-, \vec
a_\perp) = ((a^0+a^3)/\sqrt{2}, (a^0-a^3)/\sqrt{2}, a^1, a^2)$ and
$a_\perp^2 =(a^1)^2+(a^2)^2$. We take $\pi$ with the momentum $P^\mu
=(P^+, P^-,0,0)$ with $P^+$ as the large component and introduce a
vector $u^\mu=(u^+,u^-,0,0)$. The TMD light-cone wave function of
$\pi$ can be defined as in the limit $u^+ << u^-$:
\begin{eqnarray}
\phi_{+}(x, k_\perp,\zeta, \mu) &=& \ \int \frac{ d z^- }{2\pi}
  \frac {d^2 z_\perp}{(2\pi )^2}  e^{ik^+z^- - i \vec z_\perp\cdot \vec
k_\perp}
\langle 0 \vert \bar q(0) L_u^\dagger (\infty, 0)
  \gamma^+ \gamma_5 L_u (\infty,z) q(z) \vert \pi(P) \rangle\vert_{z^+=0},
\nonumber\\
   k^+ &=& xP^+, \ \ \ \ \  \zeta^2 = \frac{2 u^- (P^+)^2}{u^+}\approx
\frac{ 4 (u\cdot P)^2}{u^2},
\end{eqnarray}
where $q(x) $ is the light-quark field. We do not specify if the
light quark $q$ is a $u$- or $d$-quark and $\pi$ is $\pi^0$ or
$\pi^+$, which is not important for our discussion. $L_u$ is the gauge
link in the direction $u$:
\begin{equation}
L_u (\infty, z) = P \exp \left ( -i g_s \int_{0} ^{\infty} d\lambda
     u\cdot G (\lambda u+ z ) \right ) .
\end{equation}
It should be understood that the contributions proportional to any
positive power of $u^+/u^-$ are neglected. This definition was first
proposed in \cite{BS}. It is gauge invariant in any non-singular
gauge in which the gauge field is zero at infinite space-time. The
TMD wave function has an extra variable $\zeta^2$ beside the
momentum faction $x$, the transverse momentum $k_\perp$ and the
renormalization scale $\mu$. The evolution of this variable will
generate the resummed Sudakov logarithms as shown in \cite{BS}. This
will be confirmed in this letter. The evolution with the
renormalization scale $\mu$ is simple:
\begin{equation}
\mu \frac{\partial \phi_+(x, k_\perp,\zeta, \mu) }{\partial \mu }
= 2\gamma_q  \phi_+(x, k_\perp,\zeta, \mu),
\end{equation}
where $\gamma_q$  is the anomalous dimension of the light quark
field $q$ in the axial gauge $u\cdot G=0$. In the axial gauge,
the gauge links in our definition disappear. Some general features
of TMD light-cone wave functions defined by using the light-cone gauge link
with $u^+=0$ have been studied in
\cite{BFLS,CZ}. In \cite{BFLS} the operator product expansion was employed
to analyze the short distance behavior of wave functions. However,
the TMD light-cone wave functions defined  with $u^+=0$
will have light-cone
singularities like $1/(1-x)$, as pointed out in \cite{Col1} for TMD
parton distributions. We will also show through our calculation that
these singularities exist if one sets $u^+=0$ at the beginning.
With a finite, but large $\zeta$,  the
light-cone singularities are regularized. Because of the small but finite
$u^+$, the TMD light-cone wave function is not the wave function introduced
in the light-front Hamiltonian formulation of QCD\cite{BPP}.
However, with our definition one can still derive the so-called Drell-Yan-West
relation(See \cite{Close}), which shows that the behavior
of the TMD light-cone wave function of a hadron in the region of $x\to 1$ is related
to the structure function of the hadron in the region of $x\to 1$ and
the leading power behavior of the form factor factor of the hadron.
With the definition of TMD light-cone wave function it has already been shown
that the TMD factorization for the form factor  in the transition of $\gamma^* \pi \to \gamma$
can be consistently factorized at least at one-loop level\cite{MQW}.
It should be noted that the TMD light-cone wave function defined in Eq.(1) is real.
This can be shown by using parity- and time-reversal transformation
and with the fact that the TMD light-cone wave function depends on the vector
$u$ through $\zeta^2$.
\par
The standard light-cone wave function is defined as\cite{BL}:
\begin{equation}
\Phi_+(x, \mu) = \int \frac{ d z^- }{2\pi}
    e^{ik^+z^- }
\langle 0 \vert \bar q(0) L_n^\dagger (\infty, 0)
  \gamma^+ \gamma_5 L_n (\infty,z) q(z) \vert \pi(P)
\rangle\vert_{z^+=0,\vec z_\perp=0},
\end{equation}
where the gauge link is along the light-cone direction $n^\mu=(0,1,0,0)$.
Comparing the two definitions one would expect in the limit $u^+ \to 0$ that
the standard light-cone-wave function can be obtained by:
\begin{equation}
\Phi_+(x,\mu) = \int d^2 k_\perp \phi_{+}(x, k_\perp,\zeta, \mu),
\end{equation}
where the integration over $k^2_\perp$ is from $0$ to $\infty$.
However, this is not true. The reason is as follows: By the
generalized power counting rule \cite{JMYPC}, $\phi_{+}$ is
proportional to $k^{-2}_\perp$ as $k^2_\perp\to\infty$. Hence the
transverse momentum integral is ultraviolet (U.V.) divergent. In
Eq.(4) the integration over the transverse momentum is supplemented
with systematic U.V. subtractions and this generates a
renormalization scale dependence in $\Phi_+$.
This leads to the Efremov-Radyushkin-Brodsky-Lepage
evolution equation\cite{ERBL}. In the above
integration one may also implement an U.V. subtraction by
introducing a suitable cut-off for $k_\perp$. This is easy
to do at one-loop level, and but difficult to extend beyond one-loop. Also the
limit $u^+ \to 0$ or $\zeta \to \infty$ is nontrivial.
\par
Although it is tricky to establish the above relation, the two types
of wave functions are related to each other in other ways. If we
transform the TMD light-cone wave function into the impact space
\begin{equation}
\phi_{+}(x, b,\zeta, \mu) = \int d^2 k_\perp e^{i\vec k_\perp \cdot \vec b}
\phi_{+}(x, k_\perp,\zeta, \mu)
\end{equation}
it can be shown that a factorized relation exists between two wave
functions if $b$ is small:
\begin{equation}
\phi_{+}(x, b,\zeta, \mu) = \int_0^1 d y C_b(x,y,\zeta, b, \mu) \Phi_+
(y,\mu) + {\mathcal O}(b),
\end{equation}
where the function $C$ can be calculated in perturbative QCD and
does not have any soft divergence. When $b$ is small, it corresponds to a
large momentum scale, and hence its dependence must be calculable in
perturbation theory. $\zeta$ can be understood as a large scale, its
dependence shall also be perturbative. The factorization theorem
asserts that all nonperturbative effect in $\phi_{+}(x, b,\zeta,
\mu)$ is resided in $\Phi_+ (x,\mu)$ for small $b$. We will show
this is true at one-loop level and our result can be extended beyond
one-loop.
\par
Another interesting relation can be found if $k_\perp$ is large,
which can be generated by exchanges of gluons between partons inside
$\pi$ and these gluons are hard. Hence the behavior at large
$k_\perp$ can be studied with perturbative QCD. One expects the type
of factorization:
\begin{equation}
\phi_{+}(x, k_\perp,\zeta, \mu) =\frac{1}{k_\perp^2} \int_0^1 d y
C_\perp(x,y,\zeta, \mu) \Phi_+ (y,\mu)
+ {\mathcal O}(k^{-4}_\perp).
\end{equation}
The factor $k^{-2}_\perp$ is determined by the power counting rule in
\cite{JMYPC}.
\par
As discussed above, the two functions $C_b$ and $C_\perp$ can be
calculated in perturbative QCD. To do that, we take a partonic state
to calculate the two wave functions at one-loop order, in which we
have both infrared- and collinear divergences. We regularize
infrared divergences by taking a small gluon mass $\lambda$, and
collinear divergences by taking a small quark mass $m_q$. If the
factorization in Eq.(7) is correct, $C_b$ will have no singular
dependence on these small masses. It should be noted that our
results for the two wave functions at one-loop level will also be
useful to establish TMD- and collinear factorization theorems for
exclusive processes at one-loop. We take the partonic state $\vert
q(k_q), \bar q(k_{\bar q})\rangle$ to replace $\pi$ in the above
definitions, the parton momenta are given as
\begin{equation}
k_q^\mu =(k_q^+, k_q^-,\vec k_{q\perp}), \ \ \ \  k_{\bar q}^\mu =(k_{\bar
q}^+, k_{\bar q}^-, -\vec k_{q\perp}),
\ \ \ \ \ \ k_q^+ = x_0 P^+, \ \ \ \  k_{\bar q}^+ = (1-x_0) P^+ = \bar x_0
P^+.
\end{equation}
These partons are on-shell. At tree-level the wave functions are
trivial:
\begin{equation}
\phi_+^{(0)} (x,k_\perp,\zeta) =  \delta (x-x_0)
     \delta^2(\vec k_\perp -\vec k_{q\perp}) \phi_0, \ \ \ \ \
\Phi_+^{(0)} (x, \mu) = \delta (x-x_0) \phi_0,
\end{equation}
where $\phi_0$ is a product of spinors $\phi_0 =\bar v(k_{\bar q}) \gamma^+
\gamma_5 u(k_q) /P^+$.
We will always write a quantity $A$ as $A=A^{(0)} + A^{(1)} +\cdots$, where
$A^{(0)}$ and
$A^{(1)}$ stand for tree-level- and one-loop contribution respectively.
\par
At one-loop level, one can divide the corrections into a real part
and a virtual part. The real part comes from contributions of
Feynman diagrams given in Fig.1. The virtual part comes from
contributions of Feynman diagrams given in Fig.2, which are
proportional to the tree-level result.
\par
\begin{figure}[hbt]
\begin{center}
\includegraphics[width=11cm]{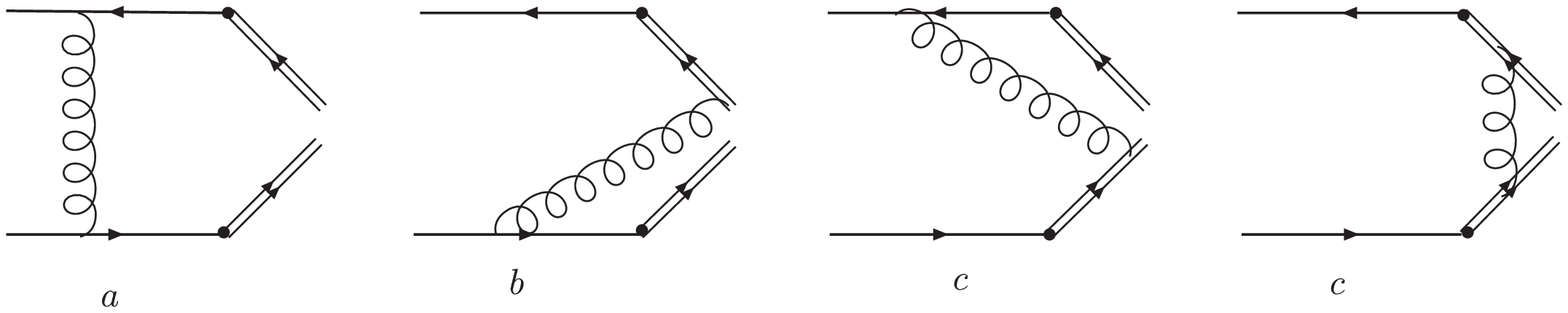}
\end{center}
\caption{The real part of one-loop contribution to the TMD light-cone wave
function.
The double lines represent the gauge links.  }
\label{Feynman-dg1}
\end{figure}
\par
The contribution from Fig.1b  and Fig.1c and Fig.1d to $\phi_+$ reads:
\begin{eqnarray}
\phi_+(x,k_\perp,\zeta)\vert_{1b} &=& -\frac{ 2 \alpha_s }{3 \pi^2}
\left \{ \theta (x_0 -x) \left ( \frac{1}{x-x_0}\right)_+ \frac{x}{x_0
\Delta_q} +
\frac{1}{2} \delta (x-x_0) \frac{1}{q_\perp^2 +\lambda^2} \ln\frac{q_\perp^2
+\lambda^2}{\zeta^2 x_0^2}
\right \} \phi_0,
\nonumber\\
\Delta_q &=& (1-y)^2 m_q^2 + y\lambda^2 + (k_\perp - y k_{q\perp})^2, \ \ \
y= \frac{x}{x_0}, \ \ \ \  q_\perp = k_\perp - k_{q\perp},
\nonumber\\
\phi_+(x,k_\perp,\zeta)\vert_{1c} &=& \frac{ 2\alpha_s }{3\pi^2}
\left \{ \theta (x -x_0) \left ( \frac{1}{x-x_0}\right)_+ \frac{1-x}{(1-x_0)
\Delta_{\bar q}}
- \frac{1}{2} \delta (x-x_0) \frac{1}{q_\perp^2 +\lambda^2}
\ln\frac{q_\perp^2 +\lambda^2}{\zeta^2 \bar x_0^2}
\right \} \phi_0,
\nonumber\\
\Delta_{\bar q}  &=& \left ( 1 -\bar y   \right )^2 m_q^2 + \bar y \lambda^2
+ \left ( k_\perp  - \bar y  k_{q\perp} \right )^2, \ \ \
\bar y = \frac{\bar x}{\bar x_0},
\nonumber\\
\phi_+(x,k_\perp,\zeta)\vert_{1d} &=& - \frac{ 2 \alpha_s  }{3\pi^2}
\frac{1}{\lambda^2 + q^2_\perp}
\delta(x-x_0) \phi_0,
\end{eqnarray}
where we have already taken the limit $\zeta\to\infty$ and only kept the
leading terms.
If we set $u^+=0$ or $\zeta =\infty$ at the beginning, the contribution from
Fig.1b and Fig.1c
will be divergent as $(x-x_0)^{-1}$ when $x\to x_0$, and Fig.1d will not
lead nonzero
contribution. The divergence is the light-cone singularity mentioned before,
caused by the
exchanged gluon
if it has the momentum $k_g^\mu =(k_g^+, k_g^-, \vec k_{g\perp})$ with a
vanishing small $k_g^+$.
With a finite, but large $\zeta$ the divergence is regularized. It results
in the $+$-distributions
and the terms with $\ln\zeta$
in the above expressions. From the above results one clearly sees that there
are collinear- and infrared singularities when transverse momenta are small.
It should be noted that the contributions
from Fig.1b and Fig.1c are related to each other by charge-conjugation.
\par
The contribution from Fig.1a
is complicated. But for the function $C_\perp$ and $C_b$  we only need
the leading part of the contribution with $k_\perp\to\infty$.
The leading part can be written as:
which is
\begin{eqnarray}
\phi_+(x,k_\perp, \zeta)\vert_{1a}  &=& \frac{2\alpha_s}{3\pi^2}
\frac{1}{ k_\perp^2 +\Lambda^2 } \left (\frac{x}{x_0} \theta (x_0-x)
    + \frac{1-x}{1-x_0} \theta (x-x_0)\right ) \phi_0
\nonumber\\
   && + "finite\ terms ",
\end{eqnarray}
where we introduce a "cut-off" $\Lambda$, which
is also contained in the second line, so that the total does not depend on
it.
The second line behaves like $k^{-4}_{\perp}$ with $k_\perp\to\infty$.
One can show that the second line is irrelevant here, but it is relevant
to what is neglected in Eq.(7,8).

\par
\begin{figure}[hbt]
\begin{center}
\includegraphics[width=8cm]{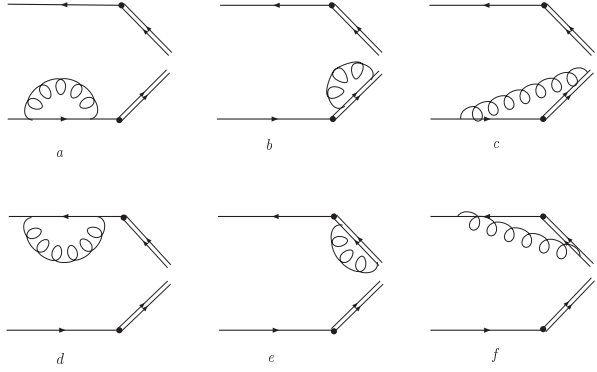}
\end{center}
\caption{The virtual part of one-loop contribution to the TMD light-cone
wave function.
The double lines represent the gauge links.  }
\label{Feynman-dg2}
\end{figure}

\par
The virtual part of the one-loop correction is from the Feynman diagrams
given in Fig.2. Contributions from each diagrams are:
\begin{eqnarray}
\phi_+ (x,k_\perp, \zeta)\vert_{2a} &=&\phi_+ (x,k_\perp, \zeta)\vert_{2d}
= \phi_+^{(0)} (x,k_\perp, \zeta) \cdot \frac{\alpha_s}{6\pi}
    \left [ -\ln \frac{\mu^2}{m_q^2} +2 \ln\frac{m_q^2}{\lambda^2} -4 \right
] ,
\nonumber\\
\phi_+ (x,k_\perp, \zeta)\vert_{2b} &=&
    \phi_+ (x,k_\perp, \zeta)\vert_{2e} = \phi_+^{(0)} (x,k_\perp, \zeta)
     \cdot \frac{\alpha_s}{3\pi}
    \ln \frac{\mu^2}{\lambda^2},
\nonumber\\
\phi_+ (x,k_\perp, \zeta)\vert_{2c} &=&
\phi_+^{(0)} (x,k_\perp, \zeta)
\cdot \frac{\alpha_s}{6\pi}
    \left [ 2\ln\frac{\mu^2}{m_q^2} + 2\ln\frac{\zeta^2 x_0^2}{m_q^2}
       +\ln^2 \frac{m_q^2}{\lambda^2} -\ln^2 \frac{\zeta^2 x_0^2}{\lambda^2}
     -\frac{2\pi^2}{3} +4 \right ],
\nonumber\\
\phi_+ (x,k_\perp, \zeta)\vert_{2f} &=&\phi_+ (x,k_\perp,
\zeta)\vert_{2c}\vert_{x_0 \to \bar x_0} .
\end{eqnarray}
These results also contain collinear- and infrared divergences.
The complete one-loop contribution $\phi_+^{(1)}$ is the sum of
contributions from the 10 Feynman
diagrams in Fig.1. and Fig.2. From the above one-loop result one can already
obtain
the relation in Eq.(8) with the function $C_\perp$ determined as:
\begin{eqnarray}
C_\perp(x,y,\zeta, \mu)& =& \frac{2\alpha_s}{3\pi^2} \left \{
      -\theta(y-x)\left [ \left ( \frac{x}{y}\frac{1}{x-y}\right )_+ -
\frac{x}{y} \right ]
      +\theta(x-y)\left [ \left ( \frac{1-x}{1- y}\frac{1}{x-y}\right )_+  +
\frac{1-x}{1-y} \right ]
\right.
\nonumber\\
  && \left. +\delta(x-y)\left [
     \ln\frac{k_\perp^2}{\zeta^2} + \ln(1-x)^2 +\ln x^2 -1 \right ] \right\}
     +{\mathcal O}(\alpha_s^2).
\end{eqnarray}
In the above the $+$-prescription acts on the distribution variable $y$. We
have used the identity
\begin{equation}
\left ( \frac{x}{y}\frac{\theta(y-x)}{x-y}\right )_+ =
\left ( \frac{x}{y}\frac{\theta(y-x)}{x-y}\right )_+  -\delta(x-y) \left
[1+\ln (1-x)\right],
\end{equation}
where the $+$-prescription in the left hand side acts on the distribution
variable $x$,
while the $+$-prescription in the right hand side acts on the distribution
variable $y$,
\par
To derive the relation in Eq.(7) one needs to transform the above results
into the impact parameter space
and to obtain one-loop results of $\Phi_+$.
The Fourier transformation can be done straightforwardly. We find
that there is a cancelation of infrared divergence between the real- and
virtual part correspondingly.
We present our results in the combination in which infrared divergences are
canceled:
\begin{eqnarray}
\phi_+(x,b,\zeta,\mu)\vert_{1b +2c} &=& \frac{2\alpha_s}{3\pi} \phi_0 \left
\{
  \frac{1}{4} \delta (x-x_0) \left [ -\ln^2 (\tilde b^2 \zeta^2 x_0^2 )
   +2 \ln (\tilde b^2 \mu^2) +2 \ln (\tilde b^2 \zeta^2 x_0^2)-4 -\pi^2
\right ]
\right.
\nonumber\\
  && \left.  + \theta (x_0-x) \left [    \frac {x}{x_0}
   \frac{\ln (\tilde b^2 m_q^2 (1- y)^2 )}{x-x_0}
   \right ]_+ \right \} +{\mathcal O}(b),
\nonumber\\
\phi_+(x,b,\zeta,\mu)\vert_{1c+2f} &=& \frac{2\alpha_s}{3\pi} \phi_0 \left
\{
  \frac{1}{4} \delta (x-x_0) \left [ -\ln^2 (\tilde b^2 \zeta^2 \bar x_0^2 )
   +2 \ln (\tilde b^2 \mu^2) +2 \ln (\tilde b^2 \zeta^2 \bar x_0^2)-4
-\pi^2\right ]
\right.
\nonumber\\
  && \left.  - \theta (x-x_0) \left [    \frac {1-x}{1-x_0}
   \frac{\ln (\tilde b^2 m_q^2 (1-\bar y)^2 )}{x-x_0}
   \right ]_+ \right \} +{\mathcal O}(b),
\nonumber\\
\phi_+(x,b,\zeta,\mu)\vert_{1d +2b+2e } &=& \frac{2\alpha_s}{3\pi} \phi_0
\delta (x-x_0)
\ln (\tilde b^2 \mu^2 )  +{\mathcal O}(b),
\end{eqnarray}
with $\tilde b = b e^\gamma/2$. The contribution from Fig.2a and Fig.2d in
the b-space
can be easily read off from Eq.(15)
With these results in the $b$-space one can derive the evolution
of $\zeta$:
\begin{eqnarray}
\zeta \frac{\partial}{\partial \zeta} \phi_+(x,b, \zeta,\mu) &=&
   \left [-\frac{2\alpha_s}{3\pi} \ln\frac{\zeta^2 x^2  b^2
e^{2\gamma-1}}{4}
        -\frac{2\alpha_s}{3\pi} \ln\frac{\zeta^2 (1-x)^2  b^2
e^{2\gamma-1}}{4}
    \right ]
       \phi_+(x,b,\zeta,\mu)+{\mathcal O}(\alpha_s^2)
\nonumber\\
       &=& -\frac{4\alpha_s}{3\pi} \left [\ln( \tilde b^2 \mu^2)
         + \ln\frac{\zeta^2 x (1-x)}{e \mu^2}
    \right ]
       \phi_+(x,b,\zeta,\mu) +{\mathcal O}(\alpha_s^2).
\end{eqnarray}
This equation was derived first in \cite{BS}. Our result agrees with that in
\cite{BS}.
The solution of the equation will contain the Sudakov logarithms in a
resummed form.
\par
The one-loop contributions to $\Phi_+$ are represented by the same
diagrams given in Fig.1 and Fig.2, except those in which the gluon
is exchanged between gauge links. Some of them contains light-cone
singularities because the gauge link is along the direction $n$,
i.e., $u^+=0$ is set at the beginning. This singularity is canceled
between the real- and virtual part correspondingly. We present our
results in the combination free from the singularity:
\begin{eqnarray}
\Phi_+ (x,\mu)\vert_{1c} +\Phi_+(x,\mu)\vert_{2f} &=&
-\frac{2\alpha_s}{3\pi} \phi_0
\theta(x-x_0) \left [ \frac{1-x}{(1-x_0)(x-x_0)} \ln \frac{m_q^2 (1-\bar
y)^2}{\mu^2} \right]_+,
\nonumber\\
\Phi_+ (x,\mu)\vert_{1b} +\Phi_+(x,\mu)\vert_{2c} &=&
\frac{2\alpha_s}{3\pi} \phi_0
\theta(x_0-x) \left [ \frac{x}{x_0(x-x_0)} \ln \frac{m_q^2 (1- y)^2}{\mu^2}
\right]_+ .
\end{eqnarray}
Again, the contribution to $\Phi_+$ from Fig.1a is complicated. However
the relevant part is just by integrating the leading part  of
$\phi_+\vert_{1a}$ over $k_\perp$ with
dimensional regularization. We have the relevant part as:
\begin{eqnarray}
\Phi_+(x,\mu)\vert_{1a} &=& \frac{2\alpha_s}{3\pi} \phi_0  \left ( \ln
\frac{\mu^2}{ \Lambda^2} -1 \right )
\left [ \frac{x}{x_0} \theta (x_0-x) +\frac{1-x}{1-x_0} \theta (x-x_0)
\right ]
\nonumber\\
   && + "finite\ terms ",
\end{eqnarray}
where the factor $-1$ comes from manipulation of $\gamma$-matrices in
$d$-dimension.
\par
With the above results we can extract the function $C_b(x,y,\zeta, b,\mu)$.
At tree-level
\begin{equation}
C_b^{(0)}(x,y,\zeta, b,\mu) =\delta(x-y).
\end{equation}
At one-loop we have:
\begin{eqnarray}
C_b^{(1)}(x,y,\zeta, b,\mu) &=&\frac{2\alpha_s}{3\pi}\left \{
    -\left ( \ln(\tilde b^2\mu^2) -1 \right ) \left [\frac{x}{y}\theta (y-x)
     + \frac{1-x}{1-y} \theta (x-y) \right ]
\right.
\nonumber\\
  && \left. + \frac{1}{4} \delta (x-y) \left [ -\ln^2 (\tilde b^2 \zeta^2
y^2 )
  -\ln^2 (\tilde b^2 \zeta^2 (1-y)^2 ) -8 -2\pi
\right. \right.
\nonumber\\
  && \left. \left.
    -2 (\ln y^2 + \ln (1-y)^2 ) \ln(\tilde b^2 \mu^2 )
    +2 \ln (\tilde b^2 \zeta^2 y^2)+ 2  \ln (\tilde b^2 \zeta^2 (1-y)^2)
\right ]
\right.
\nonumber\\
   && \left. + \theta (y-x) \left [    \frac {x}{y}
   \frac{\ln (\tilde b^2 \mu^2 )}{x-y}
   \right ]_+  - \theta (x-y) \left [    \frac {1-x}{1-y}
   \frac{\ln (\tilde b^2 \mu^2 )}{x-y}
   \right ]_+ \right \}.
\end{eqnarray}
In the above the $+$-prescription acts on the variable $y$.
Clearly, it is free from any collinear- or infrared
singularity as expected.
\par
To summarize: In this letter we have performed a study of the TMD
light-cone wave function of a $\pi$ meson which appears in TMD
factorization of exclusive processes. We have established two
factorized relations between the TMD- and the standard light-cone
wave function, the latter is relevant in collinear factorization of
exclusive processes. One relation is that the TMD light-cone wave
function can be written as a convolution of the standard one with a
perturbative coefficient function $C_\perp$ when the transverse
momentum is large. This relation is helpful for constructing models
of the TMD light-cone wave function. Another is that the TMD
light-cone wave function in the impact parameter $b$ space can be
written as a convolution of the standard one with a perturbative
coefficient function $C_b$ when $b$ is small. The function $C_b$
contains only perturbative effect and is determined at one-loop
level. This factorized relation can be extended beyond one-loop
level and it is useful for resummation of Sudakov logarithms. From
our results we confirm the result of the evolution equation of
$\zeta$, first derived in \cite{BS}. The solution of the equation
resums the large Sudakov logarithms.

\par
\vskip 5mm
\par\noindent
{\bf\large Acknowledgments}
\par
The authors thank Prof. X.D. Ji for reading the manuscript. This
work is supported by National Nature Science Foundation of P.R.
China.
\par

\end{document}